# Hardware-Accelerated Raycasting: Towards an Effective Brain MRI Visualization

A.M. Adeshina, R. Hashim, N.E.A. Khalid and Siti Z.Z. Abidin

**Abstract**—The rapid development in information technology has immensely contributed to the use of modern approaches for visualizing volumetric data. Consequently, medical volume visualization is increasingly attracting attention towards achieving an effective visualization algorithm for medical diagnosis and pre-treatment planning. Previously, research has been addressing implementation of algorithm that can visualize 2-D images into 3-D. Meanwhile, achieving such a rendering algorithm at an interactive speed and of good robustness to handle mass data still remains a challenging issue. However, in medical diagnosis, finding the exact location of brain tumor or diseases is an important step of surgery / disease management. This paper proposes a GPU-based raycasting algorithm for accurate allocation and localization of human brain abnormalities using magnetic resonance (MRI) images of normal and abnormal patients.

**Index Terms**—Brain tumor, graphic processing units, magnetic resonance imaging, volume visualization

—————————— ◆ ——————————

## 1 INTRODUCTION

Visualization is a way of projecting mental vision, image or picture that is not visible, present to the sight or an abstraction, visible to the mind [1]. Visualization could be categorized into scientific and information. Scientific visualization focuses on physical data such as meteorology, human body and earth while information visualization focuses on abstract, non-physical data such as financial data, bibliographic sources, statistical data etc [2].

The relevance of brain in human being cannot be over-emphasized. However, brain does not only exist in human being, it exists as well in other mammals but human brain is about three times larger, with over one hundred billion neurons [3]. Human brain is the center of nervous system controlling all activities of the human body from self-control, reasoning, planning to vision, with all features greatly pronounced, enlarged and developed in human beings.

Skull houses many brain slices. Though, all the tissues and organs in the human body is in 3-dimension but these slices are in 2-D when its images are taken with devices such as Computed Tomography (CT) and Magnetic Resonance Imaging (MRI). To perceive the complexity of the brain, in each of the slices that made up the skull, these slices exist in certain measured thickness with distances in-between. Studying the human tissues and organs extensively requires having the 2-D structures of the image obtained in 3-D model hence it is impossible to localize and allocate abnormalities in brain accurately using the 2-D slices from any of the image modalities such as CT and MRI.

A medical image system that could effectively serve its purpose must be able to carry out fast rendering at interactive speed and be robust enough to handle mass data. A real time 3-D system would be quite useful in localizing and allocating abnormalities in human brain.

This paper is a documentation of an ongoing research in medical volume visualization using brain MRI as a case study. The first part of this paper provides background understanding of medical volume visualization alongside with some of the very recent materials published in this domain. The next section proposes visualization of 2-D structure of the brain slices into 3-D model for the accurate localization and allocation of brain abnormalities. Raycasting algorithm is proposed to be hardware accelerated using Graphic Processing Unit (GPU) for immediate visualization of brain MRI.

## 2 BACKGROUND
### 2.1 Volumetric Data
In volume visualization, volumetric data is usually define using Cartesian grid, voxel and cells as illustrated in Fig. 1. This can be represented as typically as a set of samples $f(x,y,z,d)$ while d represents the data property at a location determined by $(x,y,z)$, which takes the form of a scalar, vector or even tension.

Scalar is the single valued parameter at each location in a dataset e.g. temperature and pressure. Vector is data existing with magnitude and direction. In 3-D, vector is usually represented as a triplet of values $(u,v,w)$, examples include particle projection, wind motion, gradient function etc. Tension is complex mathematical generalizations of vectors and matrices.

- *A.M. Adeshina is with the Faculty of Computer Science and Inforntion Technology, Universiti Tun Hussein Onn Malaysia, 86400 Parit Raja, Batu Pahat, Johor, Malaysia. E-mail: codedengineer@yahoo.com*
- *R. Hashim is with the Faculty of Computer Science and Inforntion Technology, Universiti Tun Hussein Onn Malaysia, 86400 Parit Raja, Batu Pahat, Johor, Malaysia. E-mail: radhiah@uthm.edu.my*
- *N.E.A. Khalid is with the Faculty of Computer and Mathematical Sciences, Universiti Teknologi Mara, 40450 Shah Alam, Selangor, Malaysia. E-mail: elaiza@tmsk.uitm.edu.my.*
- *Siti Z.Z. Abidin is with the Faculty of Computer and Mathematical Sciences, Universiti Teknologi Mara, 40450 Shah Alam, Selangor, Malaysia. E-mail: zaleha@tmsk.uitm.edu.my*

Two (2)-dimension is represented as X, Y-axis. This is just as flat structures in horizontal and vertical axis.

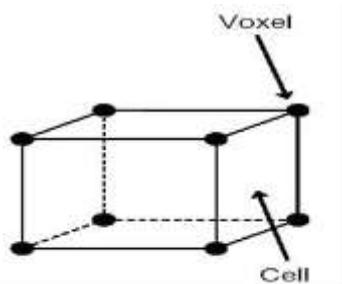

Fig. 1. Volumetric Data in Cartesian Grid

Any image in this form when turned at an angle of 90º becomes a line. Hence, a 2-dimesional structure has corners or vertices and sides in two planes. In the same vein, a 3-dimension image also has X and Y plane just as 2-D image but includes Z-axis giving the image more features and depth such as rotation. This third axis added faces to the 3-D structures.

## 2.2 Volume Visualization

Volume Visualization is a sub-field in scientific visualization that requires extraction of meaningful information from volumetric data and showing them to users via interactive computer graphics and image techniques [4].

Volume Visualization houses Surface Rendering and Direct Volume Rendering techniques as major and fundamental algorithms in volumetric data visualization.

## 2.3 Surface Rendering

Surface rendering is one of the oldest algorithms that generate primitive function, setting thresholds for original data representations in volume rendering. This is achieved using object labels as triangles, curved or flat surfaces for defining a range of voxel intensities for viewing. Surface rendering has been considered an efficient and well established approach in rendering primitives using graphic hardware but it only renders two-dimensional (2-D) surface in three-dimensional(3-D) space, losing one-dimension (1-D)[5]. Surface rendering has assisted a number of industries. Medical community renders parts of the body for diagnosis while archaeologists apply it in constructing images of specimens that are very fragile for close examination.

## 2.4 Direct Volume Rendering

Direct Volume Rendering achieves rendering by directly obtaining 2-D images of the original object from the 3-D volumetric data without any generation of intermediates. This approach actually overcomes identified drawbacks of surface rendering but with an increase in algorithm complexity and rendering time.

The processing order of the data in DVR technique is divided into three categories:

i. Object-Order DVR directly obtains 2-D images of the original object from 3-D volumetric data using forward mapping order.
ii. Image–Order DVR does the rendering the same way as Object-Oder approach above but through backward mapping. This is further separated into Maximum Intensity Projection (MIP), X-Ray Rendering and Full DVR. While MIP only writes the interpolated sample of the largest value along each ray and to each pixel, X-Ray Rendering sums all the interpolated samples along each ray and Full DVR uses sample composition of simulated light effect. Meanwhile, whenever DVR is discussed, image-order full DVR is actually referred.
iii. Domain-based DVR maps volumetric data to corresponding domain (e,g. frequency domain) before direct generation.

## 2.5 MRI Segmentation

Magnetic Resonance Imaging (MRI) is commonly employed in the medical community perhaps because of its flexibility and high spatial resolution. When brain MRI scans are produced using MRI, each point in the scan correspondence with certain points in the brain scanned. Though, there are 3-D coordinates of those points but the challenges to identify which point in the MRI scan corresponds to a particular organ or tissue in the brain is referred to as segmentation.

## 2.6 Transfer Functions

Direct Volume Rendering requires classification and rendering stages. Transfer functions are used to perform classification in DVR by assigning color and opacity to each sample in the data based on a measured property. The ability to differentiate between different tissues is of great importance to medical images. Color Transfer Function and Opacity Transfer Function are for color and opacity assignment respectively.

## 3 RELATED WORKS

3-D Visualization of volumetric medical data is an important aspect of image processing and it has shown as a promising research area due to its significance in the medical domain. The main phases in the direct volume rendering are the classification and the rendering stages. This section will discuss some of the very recent materials published in this domain.

### 3.1 Classification

Wong et al. [6] design transfer functions based on morphing factor function. The work inputs the start and the end transfer function by user for automatic generation of the intermediate transfer function.

Recently, Chu et al. [7] design an effective transfer function for direct volume rendering. The design of transfer function has a greater impact in direct volume rendering processes towards assigning color and opacity to each sample based on a measured property in the data. The work proposed a transfer function-design method based on feature variation curve. With such good design transfer function coupled with volume rendering development on GPU, the research records desirable speed and even shows that segmentation exercises might not be compulsory for volume rendering processes.

Towards the classification of data for an effective direct volume rendering, Correa & Ma [8] propose ambient ooclusion. Since transfer function does the classification in DVR, the proposed approach uses occlusion spectrum, the distribution of ambient ooclusion of a given intensity value in a 3-D volumetric data, which provides better two-dimensional  transfer functions for complex data classification. Xujia et al. [9] proposes multi-dimensional transfer function based on boundaries and present the scalar-gradient magnitude histogram.

A technique for computation of illumination using local approximation of ambient occlusion was discussed by Hernell et al. [10]. This approach avoids fully shadow region. Outcome of the work proves that a combination of multi-resolution framework and adaptive rendering with restriction of ambient ooclusion to local neighborhood produces an appreciable interactive rendering speed. In the same vein, Lindholm et al. [11] enhance DVR transfer function design with spatial localization based on user specific material dependencies. This work has contributed to encoding of knowledge of spatial relations and multi-dimensional data in transfer function (TF) domain.

As a contribution towards the finding of transfer function in Direct Volume Rendering, Liu et al. [12] propose a spreadsheet-like constructive visual component-based interface which automatically analyse histograms using the Douglas –Peucker algorithm.

Othman et al. [13] implemented Support Vector Machine (SVM) with Field Programmable Gate Array (FPGA) that offers custom computing for classification of Brain MRI which is an important phase in many of the direct volume rendering techniques. Distinguishing among patients with normal and abnormal brain MRI is the main yardstick to identify patients with brain abnormalities or tumor.

One of the earlier proposed algorithms for brain tissue classification for MRI is Partial Volume Estimation (PVE). This has earlier been implemented on Field Programmable Gate Array (FPGA). Koo et al. [14] improve on this algorithm and extended it to include probability density estimation. Human Brain MRI was used for the analysis and evaluation. Considerably better performances were achieved with the FPGA-based probability density estimation researched.

### 3.2 Rendering

Shihao et al. [15] propose a direct volume rendering with 3-D texture using hardware-assisted texture mapping.  As the outstanding challenges in medical   domain is how to render a 3-D image fast, the implementation uses trilinear interpolation to accelerate rendering speed. Some samples of engine CT scan, a human MRI head data set, beetle CT scan and hand CT scan data set were exploited.

Visualization Tool Kit from Kitware Inc. has been a great research development tool for research. Ling et al. [16] employ the use of VTK for context-preserving volume rendering. The idea was to contribute to the improvement of Maximum Intensity Projection (MIP) technique. Local Maximum Intensity Projection (LMIP) is an extended version of MIP introduced to overcome the shortcoming of MIP which is its inability to adequately depict the spatial relationships of overlapping tissues. The work presents a better and improved Local Maximum Intensity Projection that computes threshold and shading for LMIP.

With the faster and more stable Graphics Processing Units (GPUs) from Intel, Chen & Hao [17] implemented a GPU-based volume ray casting for re-sampling and representation of 3-D texture onto a sampling surface. Fragment shaders were performed with the raycasting algorithm. Visual human male CT, human head MRI and pet chest data scan were used for the experiment. The research proved an interactive speed with the algorithm for high image data.

Ray casting technique has been noted to produce high-quality images in direct volume rendering. Kim & Jaja [18] implemented cell processor architecture for broadband engine with regular datasets for direct volume rendering. Similarly, Cox et al. [19] propose parallel cell architecture broadband engine processor for speeding up the ray casting of irregular datasets. The work still requires optimization of the approach.

Direct Volume Rendering involves sampling and resampling of data. The resampling stage is carried out with the use of filters, typically trilinear, in order to ensure efficient and quality images. In some cases, quadratic or cubic filters are used for higher image quality but they are expensive to evaluate even with GPU acceleration. In addition, Csébfalvi & Domonkos [20] propose a frequency-domain upsampling on an optimal Body-Centered Cubic (BCC) lattice, which was demonstrated to have similar quality as cubic filters.

Image rendering in an interactive speed has become the greatest concern of researchers. Bentoumi et al. [21] implemented a shear-Warp algorithm on GPU. They also studied the performance in terms of image quality and rendering speed with the ray marching algorithm implemented on GPU. In order to have acceptable quality of rendered image, add in-between slices computed by interpolation was introduced to have more number of slices. However, Mensmann et al. [22] present GPU-based ray casting technique using the new programming interfaces for stream processing.  CUDA was introduced by NVIDIA as a parallel computing architecture and programming model for their GPUs and the work has shown its performance in volume rendering.

In the design of transfer functions for Direct Volume Rendering classifications, color and opacities are used to represent field values. This has introduced a limitation in using direct volume rendering mainly for higher dimensional data. In contribution to this, Manke & Wűnsche [23] introduce texture-enhanced DVR for visualization of supplementary data like material properties and additional data fields.

# 4 RESEARCH METHODOLOGY

## 4.1 Raycasting Algorithm

Raycasting is an image-order volume rendering technique firstly proposed by Levoy in 1988 (Gong & Wang 2010) [24]. The principle is based on the concept of light ray penetrating the volume in which the voxels encountered along the path would be highlighted. With respect to viewing point, a viewing ray from each of pixel on the image plane casts into the data set. The ray will make intersection with the volume dataset hence a sampling point S will be selected at an evenly spaced interval along each viewing ray. This is as illustrated in Fig. 2.

Classification in direct volume rendering is through transfer functions.

The Ray Casting process is through three (3) steps:

1. determining the path of the ray from the object to the eye position
2. determining the values encountered along the ray
3. processing the values encountered based on Ray Casting function

The following are some of the ray casting functions:

1. Maximum Intensity Projection (MIP)
2. Average Intensity Projection
3. IsoSurface
4. Thresholding
5. Compositing

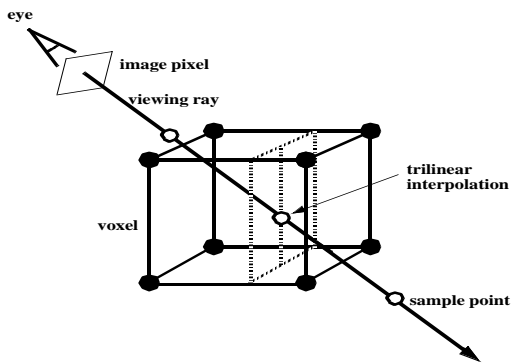

Fig. 2. Levoy proposed Ray casting Technique

Composting is proposed for this research and the next section explains the Composting Function as related to raycasting technique.

## 4.2 Compositing

Compositing is a raycasting technique that derives its final pixel color using a mix of all the sample colors and opacities encountered along the ray. The determination of each sample is achieved through transfer function by using a color and opacity. Transfer function has the greatest advantage of highlighting different tissues with different colors and opacities.

Some mathematical expressions assist in calculating ray integral in compositing. Fig. 3. gives absorption and emission along light rays paths.

Let assume a volume of density ρ (x,y,z) and Ray R coming from the back of the volume S=0 and to the eye of the viewer at $S_{eye}$. The density is expressed as:

$$\rho(x(S), y(S), z(S)) = \rho (S) \text{ -------------------------------(1)}$$

where S denotes distance.

On a normal circumstance, we expect to have density distribution inside the volume and this attenuated the light allowing only some in reaching the viewer at $S_{eye}$. If we assume to have ray path at points $S_0$, $S_1$, $S_2$, …$S_n$, we equally assume there are emissions and absorptions but no scattering at each of the points, $S_0$ to $S_n$.

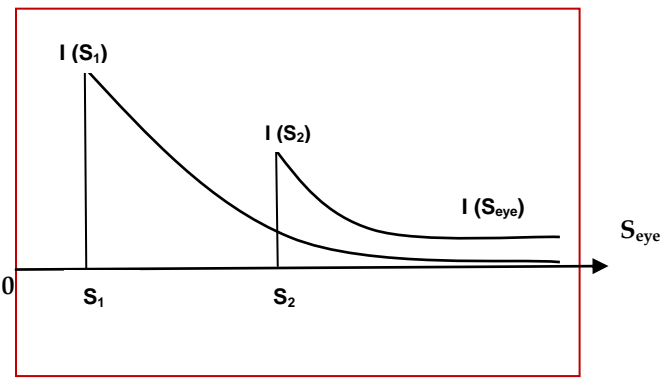

Fig. 3. Absoption and Emission along light rays

At point $S_0$, we have initial intensity at $I_0$,

$$I (S) = I (S_0) \text{ -----------------------------------------(2)}$$
I= Intensity, S= distance

If this occurs without absorption, we have all the light reaching point $S_{eye}$. Meanwhile, with absorption along $S_0$ to $S_{eye}$ and we have,

$$I (S) = I (S_0) \, e^{-\tau (S_0, S_{eye})} \text{ ----------------------------------- (3)}$$

$e^{-\tau (S_0, S_{eye})}$ is the absorption

At another point $S_1$, we have similar occurrence of emission and absorption and then we can have the intensity measured at the eye position to be:

$$I(S_{eye})=I(S_0)e^{-\tau(S_0, S_{eye})} + I (S_1)e^{-\tau (S_1, S_{eye})} + I(S_2)e^{-\tau (S_2, S_{eye})} \text{ ---(4)}$$

However, at every point along the ray path, we assume to have absorption and emission, hence, the total intensity that reaches the eye is calculated by the sum of all the intensities attenuated in all the points

$$I(S) = \int_0^s I(\check{S})\, e^{-\tau\,(\check{S},\,S)}\, dS \qquad \text{------------------(5)}$$

Where S is is the distance at eye ($S_{eye}$), $\check{S}$ is the point in the ray path (i.e. $S_0$ to $S_n$) and $\tau$ is the extinction coefficient. The above equation does the compositing in ray casting technique.

### 4.3 Visualization ToolKit

This research aims to design a 3-D visualization system with raycasting hardware-accelerated from visualization tool kit. The implementation platform is .NET, using C#.

Visualization ToolKit (VTK) [25] is an open-source visualization application with enormous software components from which visualization application can be developed. It is an object-oriented application created and distributed by kitware Inc. team with support for hundreds of algorithms, such as scalar, vector, tension texture and volumetric data.

This research considers developing the 3-D Visualization system from VTK as it enables programmatic abstraction of the access to arbitrary dimensions in the image data in comparison with other tools like trackoR, trackViz, MBAT and others [26].

This research is working towards two (2) fundamental criteria; the fast rendering at interactively speed and the 3-D visualization system's robustness towards handling mass data. Hence, the following experimens are intended for the research:

i.     Experiments with Brain MRI using CPU Ray Casting

ii.    Experiments with Brain MRI using variable capacities of NVIDIA Graphic card

Fig. 4 and Fig. 5 are the framework and research pipeline respectively.

### 4.4 Data Collection

Fifty to hundred raw volume image data of brain MRI consisting of normal and abnormal patients are intended for this research. The MRI brain images would be a combination of adult male and female skulls with age ranging between 20 to 60 years.

The reserach preliminary work commenced with representation of 2-D images (as in Fig. 6.), from which 3-D algorithm would be built.

## 5   HARDWARE REQUIREMENTS

The implementation of this research is in progress. A PC with Central Processing Unit (CPU) is the first consideration while PC with Graphic Processing Units (GPUs) of variable NVIDIA GEFORCE graphic cards capacities would be fully considered for the experiment. GPU of the following minimum specificiation is an ideal:

- Intel Core i7-950 processor (3.06Ghz/8MB/1366FSB)
- 4GB DDR3 PCI1333
- 60GB SSD Storage
- 1500GB 7200RPM Sata 2 hard disk
- 2 GB DDR3 Nvidia Geforce 430GT Graphic Card
- Windows Operating System

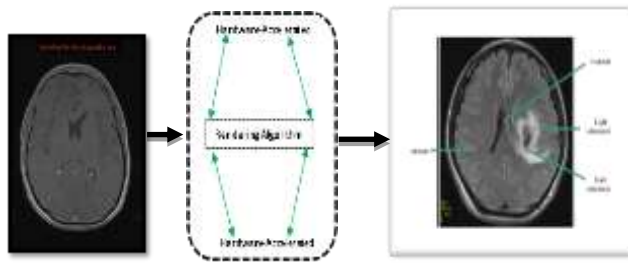

Fig. 4. Research Framework

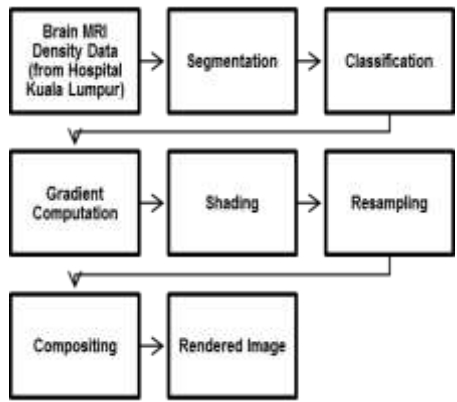

Fig. 5. Research Pipeline

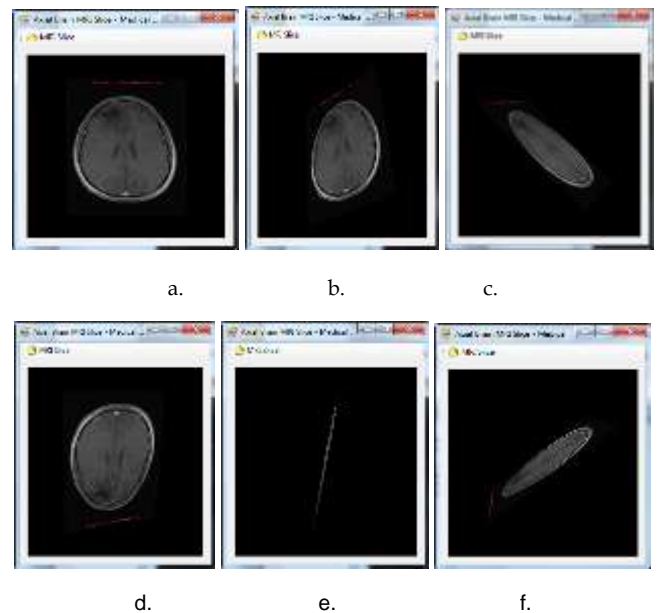

a.        b.        c.

d.        e.        f.

Fig. 6. Preliminary implementation of the hardware-accelerated Brain MRI. Fig. 6.(a-f) is the representation of 2-D images. 2-D image becomes a line when turned around as seen in Fig. 6.(e)

# 6 VTK CLASSES DESIGN

The design of vtk classes compulsorily requires the specification of a number of components. The first component is the correct specification of volume data. Raycast function is concerned with the raycasting computation integral for the final image pixel color, GPUVolumeRayCastFunction is for GPU-based, this function is used for the computations and must be specified.

VtkVolumeRayCastMapper is the vtk class that works with vtkImageData and vtkVolumeRayCastFunction. Volume data from vtkImageData are mapped by this class to calculate the pixels which will be displayed in the render window once signaled by vtkRenderer. This is refered to as the mapper for the data renderings and must be specified.

Another compulsory component that must be specified in the design of vtk class is camera settings for viewing the data.

The vtkColorTransferFunction and vtkPiecewiseFunction are transfer functions for color and scalar opacity respectively and both serves as input to vtkVolumeProperty. Fig. 7 shows typical vtk classes for volume rendering application.

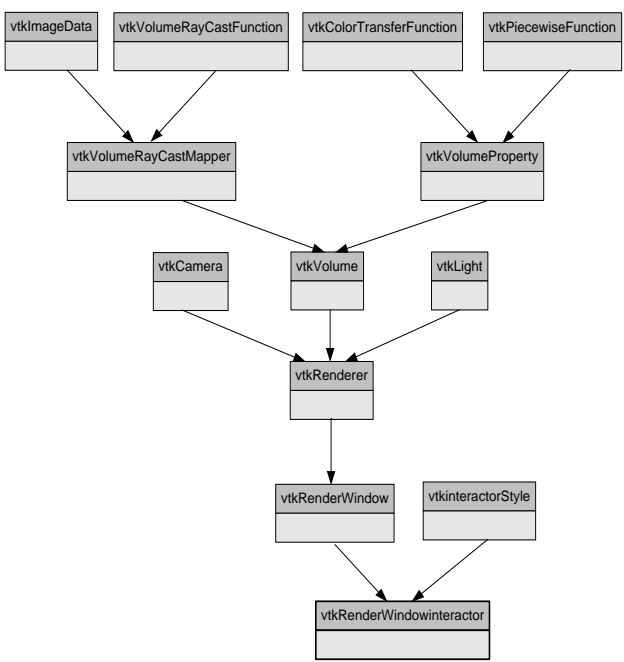

Fig. 7. Typical Vtk Classes for Volume Rendering

# 7 CONCLUSION

The existence of standard surface modeling has only assisted in defining the opaque surface of objects; this hinders us from seeing the inner part of the object. With volume visualization, we can make the boundaries of the object transparent and hence the inner part would become visible. Making the 2-D image structure of Brain MRI visible by converting it to 3-D model will facilitate localization and allocation of brain abnormalities in medical analysis. Hence, GPU accelerated raycasting for direct volume rendering of Brain MRI will be a meaningful contribution to the application of MRI data in human brain diagnosis and treatment.